# Quantum Discord and Entanglement in Quasi-Werner States Based on Bipartite Superposed Coherent States

**Ajay K Maurya[1], Manoj K Mishra[1] and Hari Prakash[1, 2]**

[1]Physics Department, University of Allahabad, Allahabad, India

[2]Indian Institute of Information Technology, Allahabad, India

Emails: *ajaymaurya.2010@gmail.com*, *manoj.qit@gmail.com*

and *prakash_hari123@rediffmail.com, hariprakash@iiita.ac.in*

**Abstract:** Recently, the authors [arXiv quant-ph 1209.3706] studied quantum discord for Werner mixed states based on the four entangled coherent states, which are used for quantum teleportation of single qubit information encoded in superposed coherent state and compared it with entanglement of formation. In the present paper, we consider two more general bipartite superposed coherent states and get two quasi-Werner states. We study quantum discord and entanglement of formation for these states and compare them.

## 1 Introduction

In 1935, Einstein *et al* [1] introduced a very surprising phenomenon called long range EPR correlation also termed as quantum entanglement [2]. At that time, it gave a new question to the researchers how one can quantify the quantum correlations present in the various types of quantum systems. However, one can say that entanglement may be considered for quantifying the quantum correlations. In some studies [3, 4], it was seen that entanglement is not unique measure of quantum correlations for any quantum system and there may be some other measures which also quantify the quantum correlations for quantum systems. In this context, Henderson *et al* [3] have shown that how one can obtain the classical part of any correlated bipartite quantum system. Authors [3] also concluded that this classical part can be used as a measure of classical correlations.

Ollivier and Zurek [4] introduced a new measure of quantum correlations called quantum discord. Quantum discord by definition is difference of the quantum versions of two classical expressions of mutual information, which are equivalent in the classical information theory. In the classical information theory, for two discrete random variables, two equivalent expressions of mutual information are given as

$$I(X:Y) = H(X) + H(Y) - H(X,Y), \qquad (1)$$

$$J(X:Y) = H(X) - H(X|Y), \qquad (2)$$

where $H(X)$, $H(Y)$ and $H(X,Y)$ are Shannon entropies [5] for random variables *X*, *Y* and the pair (*X*, *Y*) respectively, while $H(X|Y)$ is conditional entropy [5]. The quantum versions [4] of two expressions of mutual information for a given state $\rho_{X,Y}$ can be given as

$$I(X:Y) = S(\rho_X) + S(\rho_Y) - S(\rho_{X,Y}), \tag{3}$$

$$J(X:Y)_{\{\Pi_j^Y\}} = S(\rho_X) - S(\rho_{X|\{\Pi_j^Y\}}). \tag{4}$$

Here $X$ and $Y$ stand for quantum subsystems. $S(\rho_X)$, $S(\rho_Y)$ and $S(\rho_{X,Y})$ represent the von Neumann entropies [6] for the quantum subsystems $X$, $Y$ and jointly $X$, $Y$ respectively. $\{\Pi_j^Y\}$ represents the measurement basis for the subsystem $Y$. $S(\rho_{X|\{\Pi_j^Y\}})$ is conditional entropy [4] for the subsystem $X$ when the complete measurement over subsystem $Y$ is done and it can be given as

$$S(\rho_{X|\{\Pi_j^Y\}}) = \sum_j p_j S(\rho_{X|\Pi_j^Y}). \tag{5}$$

with

$$p_j = \mathrm{Tr}_{X,Y}(\Pi_j^Y \rho_{X,Y}) \text{ and } \rho_{X|\Pi_j^Y} = \frac{1}{p_j}(\Pi_j^Y \rho_{X,Y} \Pi_j^Y).$$

Quantum discord [4] is defined as

$$\begin{aligned} D(X:Y)_{\{\Pi_j^Y\}} &= I(X:Y) - J(X:Y)_{\{\Pi_j^Y\}} \\ &= S(\rho_Y) - S(\rho_{X,Y}) + S(\rho_{X|\{\Pi_j^Y\}}). \end{aligned} \tag{6}$$

Clearly quantum discord depends only on the density matrix $\rho_{X,Y}$ and the measurement basis $\{\Pi_j^Y\}$ of the subsystem $Y$. Hence one can obtain the minimized value of quantum discord over all possible measurement basis $\{\Pi_j^Y\}$ of the subsystem $Y$. This value is point of interest as it represents the quantum discord in that measurement basis in which the subsystem $X$ is least disturbed and one can extract more information about the subsystem $X$. Hence measurement basis $\{\Pi_j^Y\}$ corresponds to the basis in which conditional entropy $S(\rho_{X|\{\Pi_j^Y\}})$ attains minima. In fact quantum discord measures the amount of information that cannot be obtained by performing the measurement on one subsystem alone. Thus, the quantum discord can be considered as measure of the quantum correlations.

After this a number of studies on the quantum discord [7-12] have been done to understand the quantum correlations existing in the bipartite states. Also in a number of studies [13-20], the authors tried to quantify the quantum correlations present in the various types of bipartite systems. Further, authors [21-23] extend the study of quantum discord from bipartite to tripartite systems. Chakrabarty *et al* [22] introduced the quantum dissension for three qubits. Wang *et al* [24] have shown that non-zero quantum discord is sufficient to teleport the quantum information even when entanglement is zero.

In a recent paper [25], the authors studied comparatively the quantum discord and entanglement for quasi-Werner states based on bipartite entangled coherent states ($\sim |\alpha,\alpha\rangle \pm |-\alpha,-\alpha\rangle$ and $\sim |\alpha,-\alpha\rangle \pm |-\alpha,\alpha\rangle$). Out of the four bipartite entangled coherent states, two are maximally entangled and they form perfect Werner states, while the other two are non-maximally entangled and they form quasi-Werner states. Quasi-Werner states differ from the perfect Werner states in the sense that quantum discord and entanglement of these states depend not only on the mixing parameter but also on the mean photon number. The authors [25] also concluded that for large mean photon number, quasi-Werner states behave like perfect Werner state and then dependency of quantum discord on measurement basis disappeared.

In the present paper, we consider more general bipartite superposed coherent states (SCS), viz, $\sim |\alpha,\beta\rangle \pm |-\alpha,-\beta\rangle$ and the two modes of bipartite SCS may have different mean photon numbers. The two bipartite SCS are non-maximally entangled states and we get two quasi-Werner states based on these states. We study comparatively the dynamics of quantum discord and entanglement of formation for the two quasi-Werner states.

## 2 Quantum discord and entanglement of formation of quasi-Werner states

Let us consider the two coherent states $|\alpha\rangle$ and $|\beta\rangle$ having the mean photon numbers $|\alpha|^2$ and $|\beta|^2$ respectively. The states $|-\alpha\rangle$ and $|-\beta\rangle$ are $\pi$ radian out of phase with the corresponding coherent states. For these coherent states, we can write two bipartite SCS as

$$|\psi^\pm\rangle_{XY} = n_\pm [|\alpha,\beta\rangle \pm |-\alpha,-\beta\rangle]_{XY}, \tag{7}$$

where,

$$n_\pm = [2(1 \pm x_\alpha^2 x_\beta^2)]^{-1/2} \tag{8}$$

with

$$x_\alpha = \exp(-|\alpha|^2), \ x_\beta = \exp(-|\beta|^2).$$

One can define the even and odd coherent states, $|\pm_\alpha\rangle$ and $|\pm_\beta\rangle$ by

$$|\pm_\alpha\rangle = N_\pm^\alpha [|\alpha\rangle \pm |-\alpha\rangle],$$
$$|\pm_\beta\rangle = N_\pm^\beta [|\beta\rangle \pm |-\beta\rangle], \tag{9}$$

where $N_\pm^\alpha$ and $N_\pm^\beta$ are normalization constants given as

$$N_\pm^\alpha = [2(1 \pm x_\alpha^2)]^{-1/2},$$
$$N_\pm^\beta = [2(1 \pm x_\beta^2)]^{-1/2}. \tag{10}$$

Using equation (9), the bipartite SCS given by equation (7) can be written as

$$|\psi^+\rangle_{XY} = \frac{n_+}{2}[\frac{|+\alpha,+\beta\rangle}{N_+^\alpha N_+^\beta} + \frac{|-\alpha,-\beta\rangle}{N_-^\alpha N_-^\beta}]_{XY},$$

$$|\psi^-\rangle_{XY} = \frac{n_-}{2}[\frac{|+\alpha,-\beta\rangle}{N_+^\alpha N_-^\beta} + \frac{|-\alpha,+\beta\rangle}{N_-^\alpha N_+^\beta}]_{XY}.$$
(11)

The two bipartite SCS are non-maximally entangled states. The expressions for concurrence of the states $|\psi^+\rangle$ and $|\psi^-\rangle$ are,

$$\frac{\sqrt{(1-x_\alpha^4)(1-x_\beta^4)}}{(1+x_\alpha^2 x_\beta^2)} \text{ and } \frac{\sqrt{(1-x_\alpha^4)(1-x_\beta^4)}}{(1-x_\alpha^2 x_\beta^2)},$$

respectively. It can be easily seen that the two bipartite SCS become almost maximally entangled and mutually orthogonal for appreciable mean photon numbers in both modes. Now we write the two quasi-Werner states based on the bipartite SCS as

$$\rho(\psi^+, a) = (1-a)\frac{I}{4} + a|\psi^+\rangle\langle\psi^+|,$$

$$\rho(\psi^-, a) = (1-a)\frac{I}{4} + a|\psi^-\rangle\langle\psi^-|,$$
(12)

where $a$ is mixing parameter ranging from 0 to 1.

We can calculate the quantum discord for the quasi-Werner states given by equation (12). Density matrix for the state $\rho_{XY}(\psi^+, a)$ is

$$\rho_{XY}(\psi^+, a) = \begin{pmatrix} \frac{1}{4} + \frac{a}{4}\left(\frac{n_+^2}{(N_+^\alpha)^2(N_+^\beta)^2} - 1\right) & 0 & 0 & \frac{an_+^2}{4N_+^\alpha N_+^\beta N_-^\alpha N_-^\beta} \\ 0 & \frac{1-a}{4} & 0 & 0 \\ 0 & 0 & \frac{1-a}{4} & 0 \\ \frac{an_+^2}{4N_+^\alpha N_+^\beta N_-^\alpha N_-^\beta} & 0 & 0 & \frac{1}{4} + \frac{a}{4}\left(\frac{n_+^2}{(N_-^\alpha)^2(N_-^\beta)^2} - 1\right) \end{pmatrix}. \quad (13)$$

The four eigenvalues of this matrix are

$$\frac{(1-a)}{4}, \frac{(1-a)}{4}, \frac{(1-a)}{4}, \frac{(1+3a)}{4},$$
(14)

and the eigenvalues of reduced density matrix $\rho_Y(\psi^+, a)$ are

$$\left(\frac{1-a}{2} + \frac{an_+^2}{4(N_+^\alpha)^2(N_+^\beta)^2}\right), \left(\frac{1-a}{2} + \frac{an_+^2}{4(N_-^\alpha)^2(N_-^\beta)^2}\right).$$
(15)

In order to find the quantum discord for the state (13), one must know the conditional entropy and this can be obtained by performing complete measurement on one quantum mode. We perform complete measurement on the quantum mode $Y$ in the basis consisting of one dimensional projectors $\{\Pi_j^Y\} \equiv \{|\pi_1\rangle\langle\pi_1|, |\pi_2\rangle\langle\pi_2|\}$ with $|\pi_1\rangle = \cos\theta|+\rangle + e^{i\phi}\sin\theta|-\rangle$, $|\pi_2\rangle = \sin\theta|+\rangle - e^{i\phi}\cos\theta|-\rangle$. After the measurement in the basis $\{\Pi_j^Y\}$, we get two outcomes $\Pi_1^Y$ and $\Pi_2^Y$. For the outcomes $\Pi_1^Y$ and $\Pi_2^Y$, the corresponding states of mode $X$ are

$$\rho_{X|\Pi_1^Y} = \frac{1}{P_1}\begin{pmatrix} \frac{1}{4} + \frac{a}{4}\left(\frac{n_+^2}{(N_+^\alpha)^2(N_+^\beta)^2}\cos^2\theta - 1\right) & \frac{an_+^2}{4N_+^\alpha N_+^\beta N_-^\alpha N_-^\beta}e^{i\phi}\sin\theta\cos\theta \\ \frac{an_+^2}{4N_+^\alpha N_+^\beta N_-^\alpha N_-^\beta}e^{-i\phi}\sin\theta\cos\theta & \frac{1}{4} + \frac{a}{4}\left(\frac{n_+^2}{(N_-^\alpha)^2(N_-^\beta)^2}\sin^2\theta - 1\right) \end{pmatrix}, \quad (16)$$

and

$$\rho_{X|\Pi_2^Y} = \frac{1}{P_2}\begin{pmatrix} \frac{1}{4} + \frac{a}{4}\left(\frac{n_+^2}{(N_+^\alpha)^2(N_+^\beta)^2}\cos^2\theta - 1\right) & \frac{-an_+^2}{4N_+^\alpha N_+^\beta N_-^\alpha N_-^\beta}e^{i\phi}\sin\theta\cos\theta \\ \frac{-an_+^2}{4N_+^\alpha N_+^\beta N_-^\alpha N_-^\beta}e^{-i\phi}\sin\theta\cos\theta & \frac{1}{4} + \frac{a}{4}\left(\frac{n_+^2}{(N_-^\alpha)^2(N_-^\beta)^2}\sin^2\theta - 1\right) \end{pmatrix}, \quad (17)$$

respectively, where the probabilities $P_1$ and $P_2$ can be given as

$$P_1 = \frac{1-a}{2} + \frac{an_+^2}{4}\left(\frac{\cos^2\theta}{(N_+^\alpha)^2(N_+^\beta)^2} + \frac{\sin^2\theta}{(N_-^\alpha)^2(N_-^\beta)^2}\right),$$

$$P_2 = \frac{1-a}{2} + \frac{an_+^2}{4}\left(\frac{\cos^2\theta}{(N_-^\alpha)^2(N_-^\beta)^2} + \frac{\sin^2\theta}{(N_+^\alpha)^2(N_+^\beta)^2}\right). \quad (18)$$

The eigenvalues of density matrix $\rho_{X|\Pi_1^Y}$ are

$$\left(\frac{1-a}{4P_1}\right), \quad \left(1 - \frac{1-a}{4P_1}\right), \quad (19)$$

while eigenvalues of density matrix $\rho_{X|\Pi_2^Y}$ are

$$\left(\frac{1-a}{4P_2}\right), \quad \left(1 - \frac{1-a}{4P_2}\right). \quad (20)$$

Using equations (14, 15, 18-20, 6), one can obtain the expression of quantum discord for the state $\rho_{XY}(\psi^+, a)$ as

$$D(X:Y)_{\{\Pi_j^Y\}} = -\left(\frac{1-a}{2} + \frac{an_+^2}{4(N_+^\alpha)^2(N_+^\beta)^2}\right)\log_2\left(\frac{1-a}{2} + \frac{an_+^2}{4(N_+^\alpha)^2(N_+^\beta)^2}\right)$$

$$-\left(\frac{1-a}{2} + \frac{an_+^2}{4(N_-^\alpha)^2(N_-^\beta)^2}\right)\log_2\left(\frac{1-a}{2} + \frac{an_+^2}{4(N_-^\alpha)^2(N_-^\beta)^2}\right)$$

$$+3\left(\frac{1-a}{4}\right)\log_2\left(\frac{1-a}{4}\right) + \left(\frac{1+3a}{4}\right)\log_2\left(\frac{1+3a}{4}\right) \quad (21)$$

$$-\sum_{j=1,2}\left(\frac{1-a}{4}\right)\log_2\left(\frac{1-a}{4P_j}\right) + \left(P_j - \frac{1-a}{4}\right)\log_2\left(1 - \frac{1-a}{4P_j}\right),$$

where $P_{j=1,2}$ are given by equation (18).

Similarly we obtain the expression of quantum discord for the quasi-Werner state $\rho_{XY}(\psi^-,a)$, which can be given as

$$D(X:Y)_{\{\Pi_j^Y\}} = -\left(\frac{1-a}{2} + \frac{an_-^2}{4(N_+^\alpha)^2(N_-^\beta)^2}\right)\log_2\left(\frac{1-a}{2} + \frac{an_-^2}{4(N_+^\alpha)^2(N_-^\beta)^2}\right)$$

$$-\left(\frac{1-a}{2} + \frac{an_-^2}{4(N_-^\alpha)^2(N_+^\beta)^2}\right)\log_2\left(\frac{1-a}{2} + \frac{an_-^2}{4(N_-^\alpha)^2(N_+^\beta)^2}\right)$$

$$+3\left(\frac{1-a}{4}\right)\log_2\left(\frac{1-a}{4}\right) + \left(\frac{1+3a}{4}\right)\log_2\left(\frac{1+3a}{4}\right) \quad (22)$$

$$-\sum_{j=1,2}\left(\frac{1-a}{4}\right)\log_2\left(\frac{1-a}{4P_j}\right) + \left(P_j - \frac{1-a}{4}\right)\log_2\left(1 - \frac{1-a}{4P_j}\right),$$

where the probabilities $P_1$ and $P_2$ can be given as

$$P_1 = \frac{1-a}{2} + \frac{an_-^2}{4}\left(\frac{\cos^2\theta}{(N_+^\alpha)^2(N_-^\beta)^2} + \frac{\sin^2\theta}{(N_-^\alpha)^2(N_+^\beta)^2}\right),$$

$$P_2 = \frac{1-a}{2} + \frac{an_-^2}{4}\left(\frac{\cos^2\theta}{(N_-^\alpha)^2(N_+^\beta)^2} + \frac{\sin^2\theta}{(N_+^\alpha)^2(N_-^\beta)^2}\right). \quad (23)$$

We conclude from equations (21) and (22) that quantum discord of the two quasi-Werner states based on two bipartite SCS depend on the mixing parameter $a$, measurement parameter $\theta$ and mean photon numbers of both modes. We plot the quantum discord against the mixing parameter $a$ and measurement parameter $\theta$ for different value of mean photon numbers of both modes. Figure **1** shows the variation of quantum discord for quasi-Werner state $\rho_{XY}(\psi^+,a)$, while figure **2** shows the variation of quantum discord for quasi-Werner state $\rho_{XY}(\psi^-,a)$.

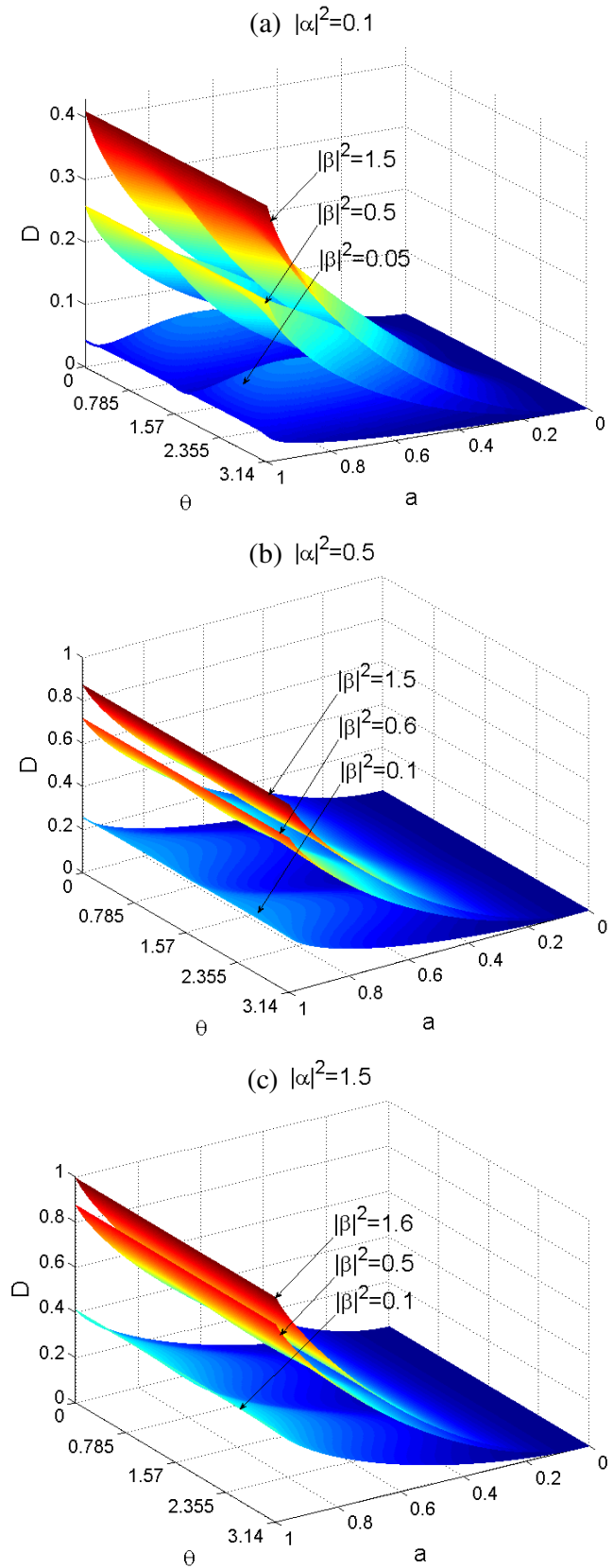

**Figure 1 (a)-(c):** Quantum discord for quasi-Werner state $\rho_{XY}(\psi^+, a)$.

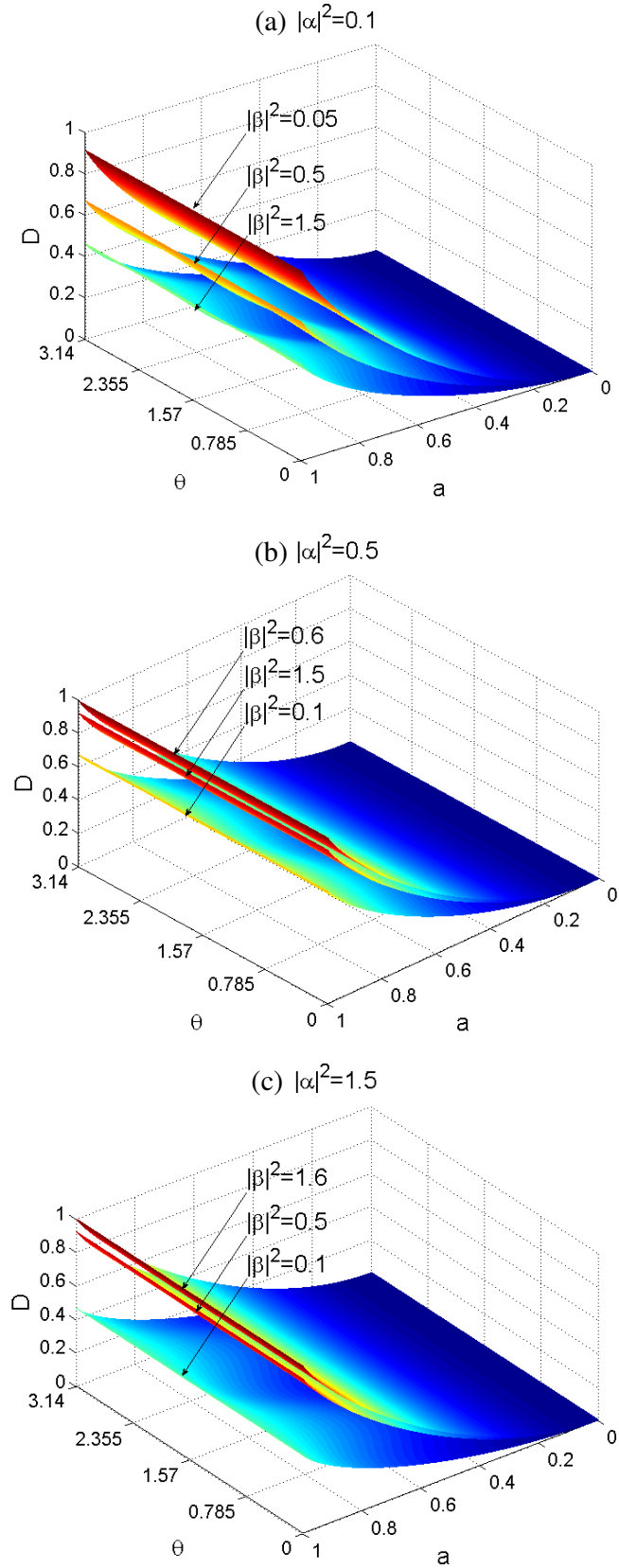

**Figure 2 (a)-(c):** Quantum discord for quasi-Werner state $\rho_{XY}(\psi^-, a)$.

From figure **1**, it is clear that quantum discord for quasi-Werner state $\rho_{XY}(\psi^+, a)$ depends on the mixing parameter $a$ and measurement parameter $\theta$ for the lower values of mean photon numbers in the both modes and quantum discord varies periodically with measurement parameter $\theta$. As the value of mean photon number in both modes increases, the periodic variations in quantum discord with measurement parameter $\theta$ tend to vanish. For appreciable mean photon number in both modes, the periodic variations in quantum discord vanish and the quantum discord for this quasi-Werner state is the same as the corresponding perfect Werner state, *i.e.*, in this situation, quantum discord depends on the mixing parameter $a$ only. Also, we see that when mean photon numbers in both modes are same, *i.e.*, $|\alpha|^2 = |\beta|^2$, we get the same results for quantum discord for these states, which are reported by the authors [25] for the corresponding quasi-Werner states.

We consider the quantum discord of the quasi-Werner state $\rho_{XY}(\psi^-, a)$. For this state, we find interesting and very different results from the quasi-Werner state $\rho_{XY}(\psi^+, a)$. From figure **2**, we find that for higher values ($\geq 1.5$) of mean photon numbers in both modes, the quantum discord becomes independent of measurement parameter $\theta$. But quantum discord varies periodically with respect to the measurement parameter $\theta$, whenever mean photon numbers in both modes are different. Thus, quantum discord depends on the mixing parameter $a$ as well as the mean photon numbers in both modes for the quasi-Werner state $\rho_{XY}(\psi^-, a)$. From the plots in figure **2**, we conclude that quantum discord attains its maximum value whenever the difference of mean photon numbers in both modes vanishes. We also conclude that as the difference of mean photon numbers increases, the value of quantum discord decreases for this state, *i.e.*, quantum correlation between two modes decreases.

Now we find the entanglement of formation for the two quasi-Werner states (equation (12)). For the quasi-Werner state $\rho_{XY}(\psi^+, a)$, square roots of the eigenvalues of the matrix $\rho_{XY}\tilde{\rho}_{XY}$ with $\tilde{\rho}_{XY} = (\sigma_y \otimes \sigma_y)\rho_{XY}^*(\sigma_y \otimes \sigma_y)$ are

$$\frac{1-a}{4}, \frac{1-a}{4}, \left[\left\{\left(\frac{1-a}{4} + \frac{an_+^2}{4(N_+^\alpha)^2(N_+^\beta)^2}\right)\left(\frac{1-a}{4} + \frac{an_+^2}{4(N_-^\alpha)^2(N_-^\beta)^2}\right)\right\}^{1/2} \pm \frac{an_+^2}{4N_+^\alpha N_+^\beta N_-^\alpha N_-^\beta}\right],$$

and for the other quasi-Werner state $\rho_{XY}(\psi^-, a)$, square roots of the eigenvalues of the matrix $\rho_{XY}\tilde{\rho}_{XY}$ are

$$\frac{1-a}{4}, \frac{1-a}{4}, \left[\left\{\left(\frac{1-a}{4} + \frac{an_-^2}{4(N_+^\alpha)^2(N_-^\beta)^2}\right)\left(\frac{1-a}{4} + \frac{an_-^2}{4(N_-^\alpha)^2(N_+^\beta)^2}\right)\right\}^{1/2} \pm \frac{an_-^2}{4N_+^\alpha N_+^\beta N_-^\alpha N_-^\beta}\right].$$

We see that for the two states, third eigenvalue is the largest. The concurrence can be obtained by the expression,

$$C(\rho_{XY}) = \max\{0, (2\lambda_{\text{largest}} - \text{Sum of the four eigenvalues})\}. \qquad (24)$$

Entanglement of formation can be obtained by the relation,

$$E(\rho_{XY}) = H(x) = -x\log x - (1-x)\log(1-x), \qquad (25)$$

where $H$ is the Shannon entropy function and $x = \frac{1}{2}(1+\sqrt{1-C^2})$.

Figure **3** shows the variation of entanglement of formation for the quasi-Werner state $\rho_{XY}(\psi^+, a)$ with respect to mixing parameter $a$ and mean photon numbers in the two modes, while figure **4** shows the variation of entanglement of formation for the quasi-Werner state $\rho_{XY}(\psi^-, a)$.

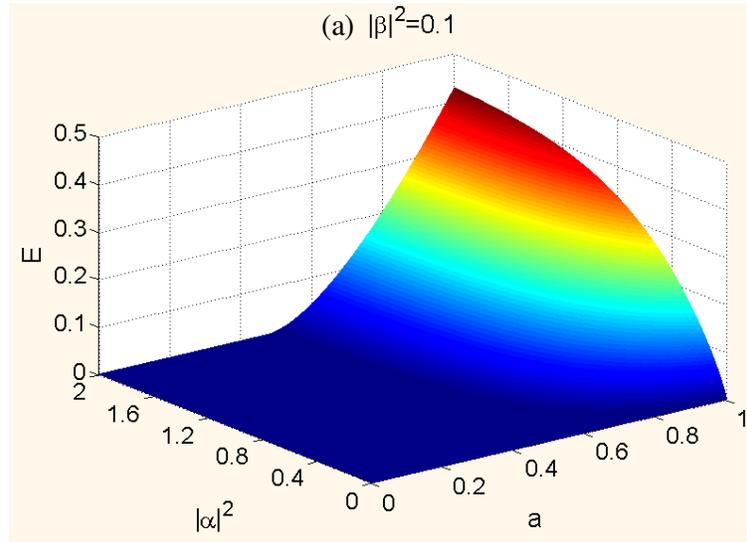

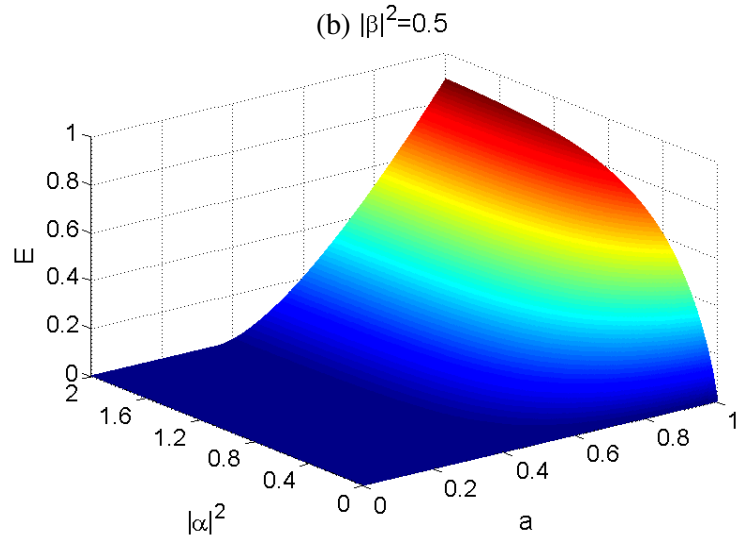

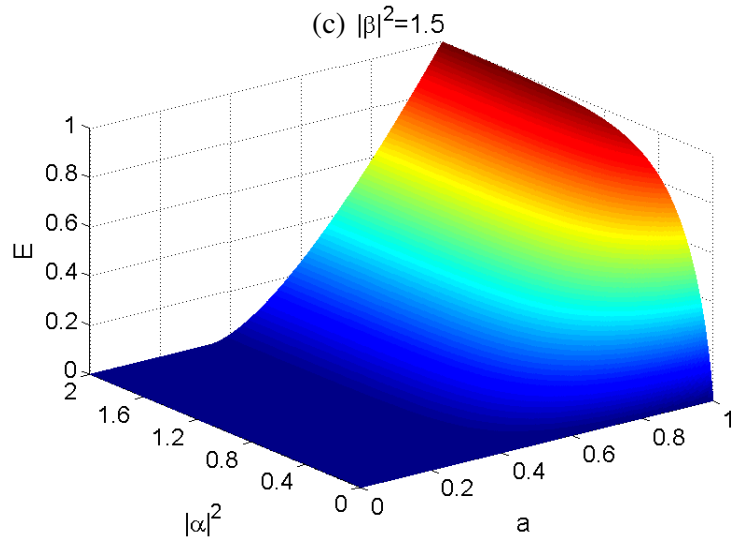

**Figure 3 (a)-(c):** Entanglement of formation for quasi-Werner state $\rho_{XY}(\psi^+, a)$.

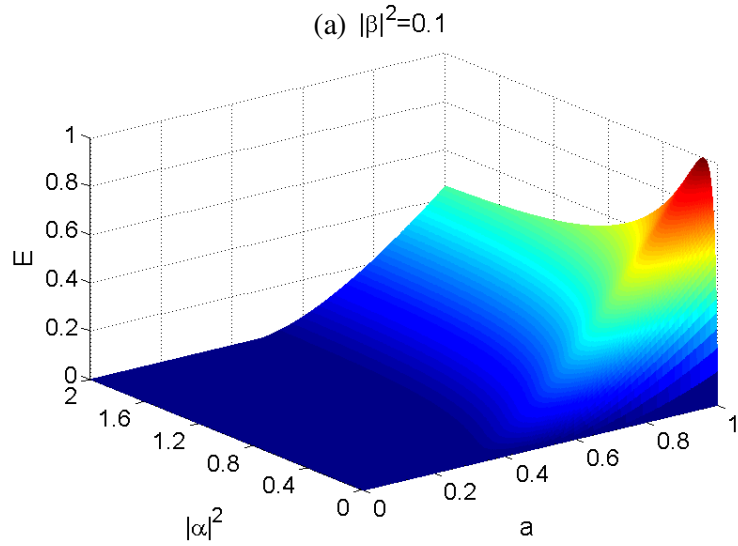

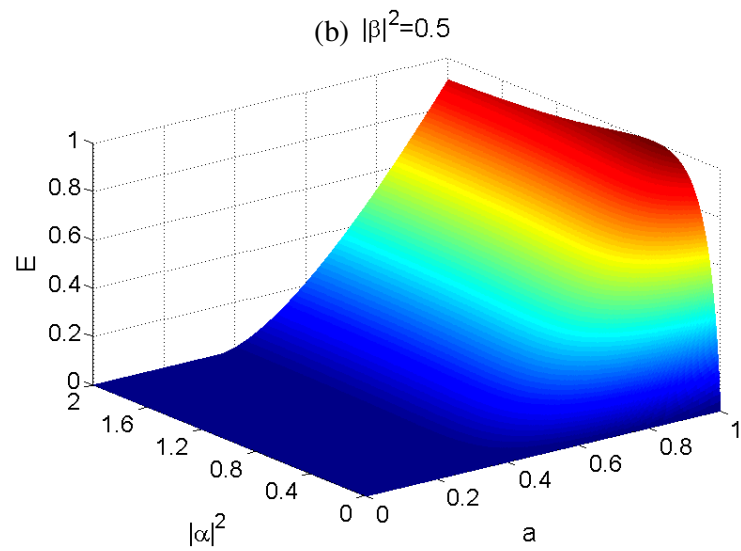

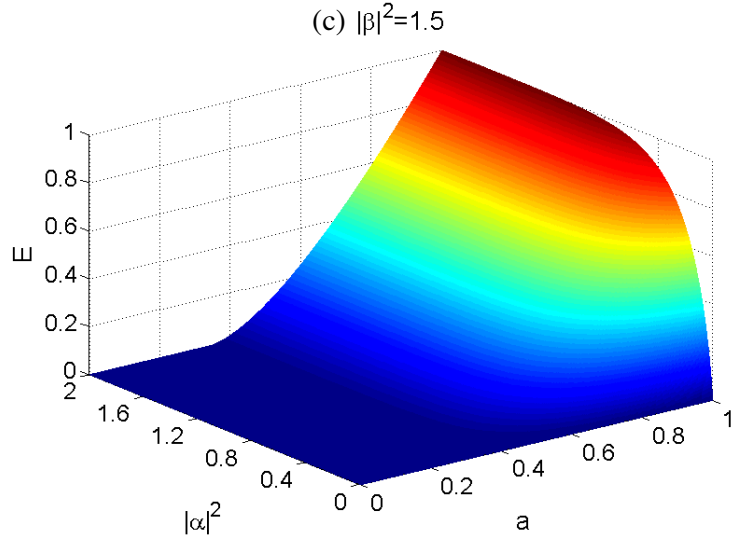

**Figure 4 (a)-(c):** Entanglement of formation for quasi-Werner state $\rho_{XY}(\psi^-, a)$.

From figure **3**, we see that entanglement of formation for the quasi-Werner state $\rho_{XY}(\psi^+, a)$ varies with respect to mixing parameter *a* as well as the mean photon numbers in the two modes and it increases with increasing mixing parameter *a* and also the mean photon numbers in the two modes.

From figure **4**, it is clear that entanglement of formation for the quasi-Werner state $\rho_{XY}(\psi^-, a)$ also varies with respect to mixing parameter *a* and the mean photon numbers in the two modes, but it attains maximum value, whenever the mixing parameter *a* is maximum and the mean photon numbers in the two modes is the same. Further, as the difference in mean photon numbers in the two modes increases, entanglement of formation decreases.

As the quantum discord is obtained by performing a complete measurement on the mode *Y*, we can consider the measurement basis for the mode *Y* for which we get the minimum value of the quantum discord. We denote this minimum value of quantum discord by $\delta$ and we can see the variation of $\delta$ with respect to mixing parameter and mean photon number in any one mode by fixing the mean photon number in other mode. Keeping this in mind, in figure **5**, we plot the difference of minimum quantum discord and entanglement of formation $\delta - E$ for the state $\rho_{XY}(\psi^+, a)$ with respect to mixing parameter and mean photon number in the mode *X*, while in figure **6**, we plot the difference of minimum quantum discord and entanglement of formation $\delta - E$ for the state $\rho_{XY}(\psi^-, a)$.

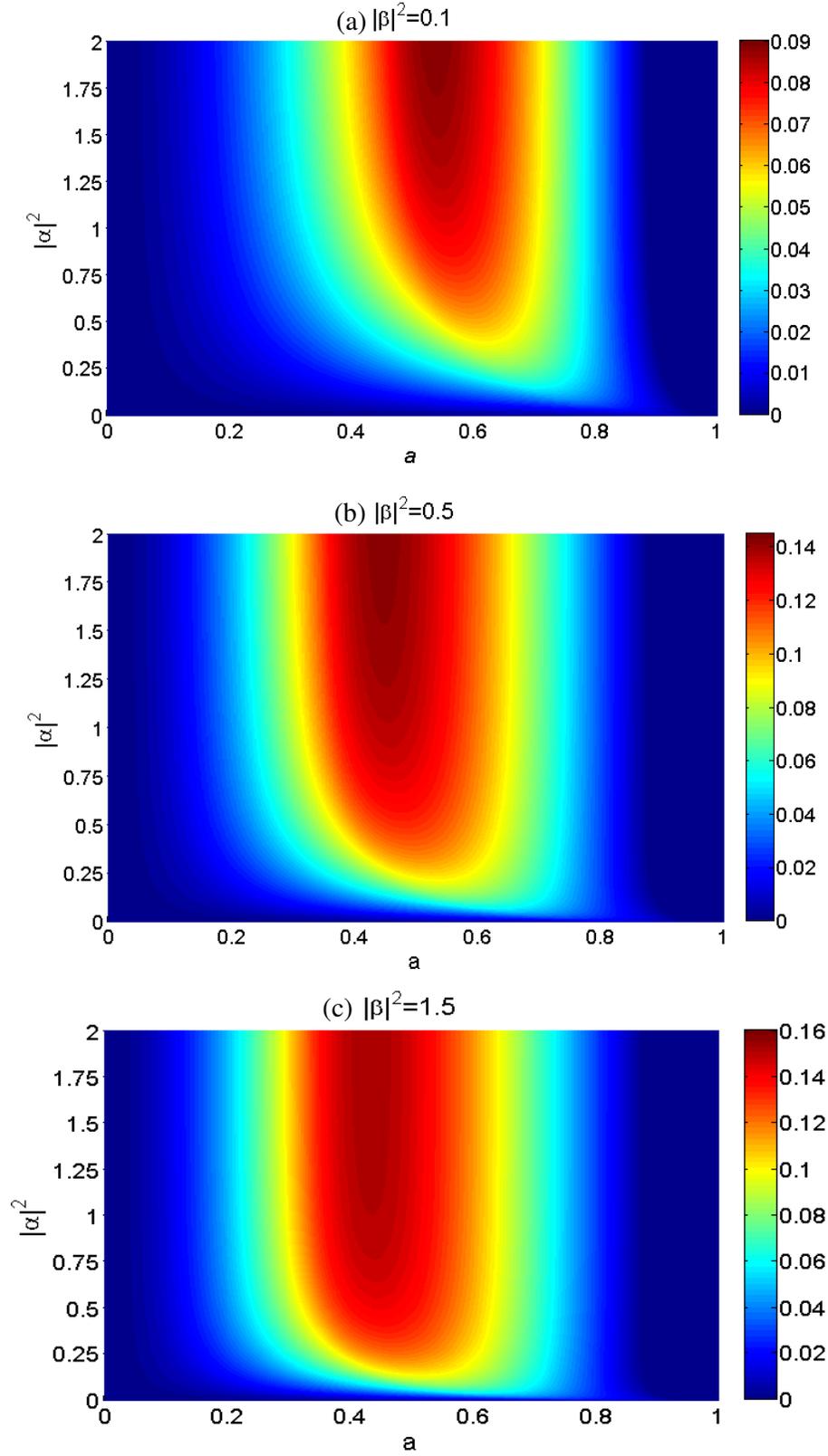

**Figure 5 (a)-(c):** Difference of minimum Quantum discord and Entanglement of formation $\delta - E$ for the quasi-Werner state $\rho_{XY}(\psi^+, a)$.

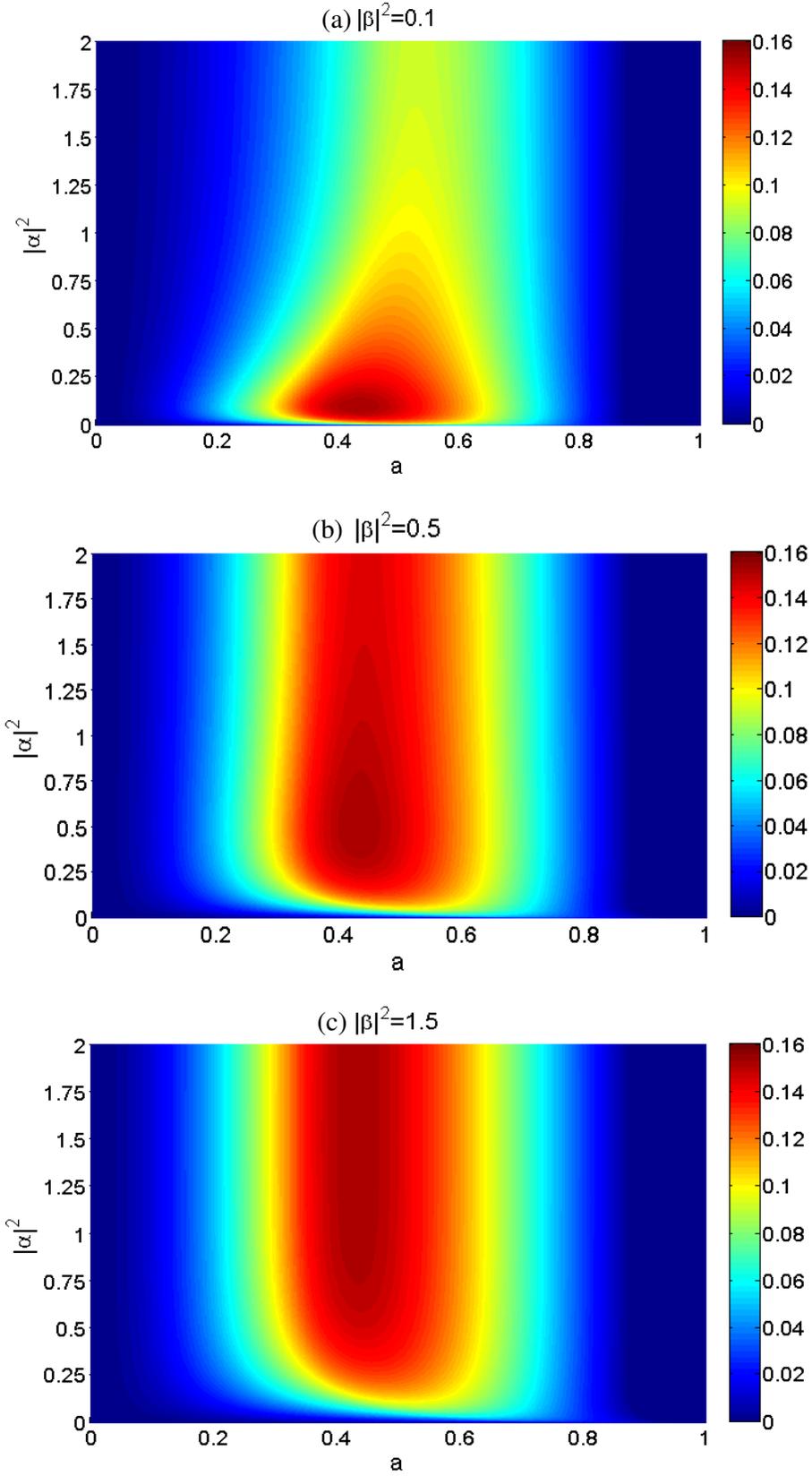

**Figure 6 (a)-(c):** Difference of minimum Quantum discord and Entanglement of formation $\delta - E$ for the quasi-Werner state $\rho_{XY}(\psi^-, a)$.

For the quasi-Werner state $\rho_{XY}(\psi^+,a)$, we can see from figure **5** that quantum discord is greater than entanglement of formation for all mixing parameter *a* except *a* =1. Also rate of increase in the value of quantum discord is higher than rate of increase in the value of entanglement of formation for the range of mixing 0 to nearly 0.4. Above this mixing range, rate of increase in the value of entanglement of formation becomes higher, but both attains the maximum value at *a* =1.

From figure **6**, we see that for the quasi-Werner state $\rho_{XY}(\psi^-,a)$, we obtain the same results which are discussed in previous paragraph for the state $\rho_{XY}(\psi^+,a)$. Also we conclude that the difference between $\delta$ and *E* is maximum when mean photon number in both modes are equal.

## 3 Conclusions

In this paper, we have shown the dynamics of quantum discord and entanglement of formation for the two quasi-Werner states based on the two bipartite SCS. For the two quasi-Werner states, quantum discord and entanglement of formation depend on the mixing and mean photon numbers in both modes. For the quasi-Werner state $\rho_{XY}(\psi^+,a)$, as we increase the mean photon numbers in both modes, both quantum discord and entanglement of formation increase. But for the quasi-Werner state $\rho_{XY}(\psi^-,a)$, both quantum discord and entanglement of formation become higher when the difference in mean photon numbers in the two modes is smaller. We also conclude that for both states, as mixing parameter *a* increases, both quantum discord and entanglement of formation increase.

For the two quasi-Werner state, we also conclude that quantum discord is greater than entanglement of formation for all mixing. The rate of increase in quantum discord is higher than rate of increase in entanglement of formation for *a* less than 0.4. After this, rate of increase in entanglement of formation dominates and at *a* =1, both attain maxima.

**Acknowledgements**

We are thankful to *Prof. N. Chandra* and *Prof. R. Prakash* for their interest in this work. We would like to thank *Dr. D. K. Singh, Dr. D. K. Mishra, Dr. R. Kumar, Dr. P. Kumar*, *Mr. Ajay K. Yadav* and *Mr. Vikram Verma* for helpful and stimulating discussions. Author AKM acknowledges financial support of CSIR under CSIR-SRF scheme and author MKM acknowledges financial support of UGC under UGC-SRF fellowship scheme.